\documentclass[12pt]{article} 
\usepackage[sectionbib]{natbib}
\usepackage{array,epsfig,fancyhdr,rotating}
\usepackage[]{hyperref}  
\usepackage{sectsty, secdot}
\sectionfont{\fontsize{12}{14pt plus.8pt minus .6pt}\selectfont}
\renewcommand{\theequation}{\thesection\arabic{equation}}
\subsectionfont{\fontsize{12}{14pt plus.8pt minus .6pt}\selectfont}

\usepackage{geometry}
\usepackage{amsmath}
\usepackage{amssymb}
\usepackage{amsfonts}
\usepackage{multirow}
\usepackage{amsthm}
\usepackage{threeparttable}
\usepackage{booktabs}

\newtheorem{prop}{Proposition}
\theoremstyle{definition}

\newtheorem{example}{Example}

\newtheorem{result}{Result}

\newtheorem{assumption}{Assumption}
\def\T{{ \mathrm{\scriptscriptstyle T} }}

\begin{document}


\renewcommand{\baselinestretch}{2}

\markright{ \hbox{\footnotesize\rm Statistica Sinica
}\hfill\\[-13pt]
\hbox{\footnotesize\rm
}\hfill }

\markboth{\hfill{\footnotesize\rm Baoluo Sun, Wang Miao and Deshanee S. Wickramarachchi} \hfill}
{\hfill {\footnotesize\rm Adjusting for Missing Data Using IVs} \hfill}

\renewcommand{\thefootnote}{}
$\ $\par


\fontsize{12}{14pt plus.8pt minus .6pt}\selectfont \vspace{0.8pc}
\centerline{\large\bf On Doubly Robust Estimation with Nonignorable }
\vspace{2pt} 
\centerline{\large\bf  Missing Data Using Instrumental Variables}
\vspace{.4cm} 
\centerline{Baoluo Sun$ ^1$, Wang Miao$ ^2 $, and  Deshanee S. Wickramarachchi$ ^1 $} 
\vspace{.4cm} 
\centerline{\it $ ^1 $Department of Statistics and Data Science, National University of Singapore}
\centerline{\it $ ^2 $Department of Probability and Statistics, Peking University}
 \vspace{.55cm} \fontsize{9}{11.5pt plus.8pt minus.6pt}\selectfont


\begin{quotation}
\noindent {\it Abstract:}
Suppose we are interested in the mean of an outcome that is subject to nonignorable nonresponse. This paper develops new semiparametric estimation methods with  instrumental variables which affect nonresponse, but not the outcome. The proposed estimators  remain consistent and asymptotically normal even under partial model misspecifications for two variation independent nuisance components. We evaluate the performance of the proposed estimators via a simulation study, and apply them in  adjusting for missing data induced by HIV testing refusal in the evaluation of HIV seroprevalence in Mochudi, Botswana, using interviewer experience as an instrumental variable.

\vspace{9pt}
\par
\end{quotation}\par

\def\thefigure{\arabic{figure}}
\def\thetable{\arabic{table}}

\renewcommand{\theequation}{\thesection.\arabic{equation}}

\fontsize{12}{14pt plus.8pt minus .6pt}\selectfont

\sloppy

\section{ Introduction}
\label{sec: intro}

Missing data are ubiquitous in the health and social sciences.  Our motivating example concerns a household survey in Mochudi, Botswana to estimate HIV seroprevalence among adults. About $19\%$ of the adults who were contacted for
the survey have missing final HIV status, mainly due to refusal to participate in the HIV testing component. The nonresponse is said to be ignorable if, conditional on
the fully observed variables, it is independent of the underlying HIV status \citep{rubin1976inference,rubin2019statistical}. In this case, the HIV seroprevalence among respondents is representative of the overall HIV  seroprevalence in the population, within strata of the observed variables. Nonetheless, ignorability is a strong assumption which may be untenable in practice;  for instance, HIV testing refusal may be entangled with features of the underlying HIV status in the household survey. The problem of nonignorable nonresponse has therefore received much attention in the missing data literature. \cite{10.1007/978-1-4612-1284-3_1} described a general class of models which requires {\it a priori} specification of a selection bias parameter that encodes the residual association of the nonresponse mechanism with the outcome of interest, after adjusting for the fully observed variables.  It coincides with the widely adopted exponential tilting model in the special case where this residual association is captured on the log odds ratio scale \citep{vansteelandt2007estimation,kim2011semiparametric,miao2015identification,miao2016varieties,shao2016semiparametric}. When the selection bias or tilting parameter in the exponential tilting model  is {\it a priori} known or can be estimated from external data, functionals of the complete data distribution such as the outcome mean can be estimated based on either the inverse propensity weighting or regression approach \citep{scharfstein1999adjusting,vansteelandt2007estimation,kim2011semiparametric}. These two approaches can be carefully combined to obtain doubly robust estimators which remain consistent and asymptotically normal if either the propensity score or regression model, but
not necessarily both, is correctly specified. Such methods have grown in popularity in recent years for estimation with missing data and other forms of coarsening, as they  effectively double one’s chances to obtain valid inference  \citep{robins1994estimation,scharfstein1999adjusting, robins2001comment, laan2003unified, bang2005doubly, tsiatis2007semiparametric,molenberghs2014handbook, seaman2018introduction,chernozhukov2022locally}. Furthermore,  these methods are also locally semiparametric efficient because they are able to achieve the
semiparametric efficiency bound when all the model specifications hold.

Doubly robust methods when the tilting parameter is unknown are far less developed and  generally require specific, non-trivial constructions.  \cite{miao2016varieties} and \cite{miao2015identification}  developed doubly robust inference  by leveraging a shadow variable which affects the outcome but not the nonresponse. Such a variable may be available in many empirical studies, and has played a prominent role in semiparametric estimation with nonignorable nonresponse \citep{liang2000regression,tang2003analysis,chen2009identifiability,d2010new,wang2014instrumental,zhao2015semiparametric,shao2016semiparametric,zhao2022versatile,limiao2023}. Another approach involves instrumental variables which affect nonresponse, but not the outcome. The  instrumental variable approach has a longstanding tradition in econometrics \citep{heckman1974shadow,heckman1979sample,manski1990nonparametric,ahn1993semiparametric,powell1994estimation, das2003nonparametric}, and has witnessed renewed interests in the health and social sciences by leveraging  certain operational features of a study \citep{BLCP:CN042547736,doi:10.1177/0049124103262689,nicoletti2005survey, barnighausen2011correcting,tchetgen2017general,sun2018semiparametric,marden2018implementation}. In the motivating example, interviewers are  randomly deployed prior to the survey,  possibly given the values of  baseline covariates  such as administrative regions. Therefore, interviewer characteristics such as years of experience constitute candidate instruments which likely influence the response rates of individuals  contacted for the survey, but are  independent of the individuals' underlying outcomes of interest, within strata of the baseline covariates.

\citet{sun2018semiparametric} established the semiparametric efficiency theory for estimating the outcome mean with instrumental variables when the tilting parameter is unknown. They also proposed a doubly robust estimator of the outcome mean  based on the widely adopted odds ratio factorization of the complete data distribution, in which the outcome density among nonrespondents can be expressed as an exponential tilting of the density among respondents \citep{robins2001comment,vansteelandt2007estimation,kim2011semiparametric,miao2015identification,miao2016varieties,shao2016semiparametric, riddles2016propensity, malinsky2022semiparametric}; see  \cite{kim2021statistical} for a recent review. However, as we show later in the paper,  doubly robust inference with instrumental variables is complicated by the fact that, under this widely adopted factorization,  the   models for different components of the complete data distribution are in fact variationally dependent. Therefore, any choice of one model imposes {\it a priori} restrictions on the range of possible models for the remaining components, which implies that we may not have two independent opportunities for valid inference. Furthermore, one may in fact rule out the possibility of achieving local semiparametric efficiency due  to possible lack of model compatibility, which  leaves a gap in the nonignorable missing data literature.   The main contribution of our paper is to provide explicit construction of novel doubly robust and locally efficient estimators which resolves these issues through an alternative factorization of the complete data distribution that naturally encodes the instrumental variable assumptions.

\section{Preliminaries}

\label{sec:prelim}

{
\subsection{Model and assumptions}

The full data $W=(Y,Z,U)$ has support $\mathcal{W}=(\mathcal{Y}\times \mathcal{Z}\times \mathcal{U})$. Here $Y$ is an outcome of interest, $Z$ an instrumental variable, and $U=(U_1,...,U_L)$ consists of $L$ measured baseline covariates. Suppose that $X=(Z,U)$ is fully observed, but $Y$ is subject to missingness. Let $R\in\{0,1\}$ denote the binary random variable indicating missingness status, with $R=1$ if $Y$ is observed and $R=0$ otherwise. We assume that the distribution of the complete data $(R,W)$ has density $$p(r,w)=\pi(w)^r\{1-\pi(w)\}^{1-r}p(w),$$ with respect to some appropriate dominating measure, where $\pi(w)=p(R=1\mid w)$ denotes the extended propensity score which captures the nonresponse mechanism, and $p(w)$ is the full data density. Throughout, we make the following positivity assumption, which is necessary for identification of the complete data distribution and smooth functionals of the latter, and ensures finite asymptotic variance of the proposed estimators \citep{rotnitzky1998semiparametric}.
\begin{assumption}
\label{assp:pos}
$\pi(w)>\sigma>0$ for all $w\in \mathcal{W}$,  where $\sigma$ is a fixed positive constant.
\end{assumption}

 Let $p(r,w;\varphi)$ denote a model for the complete data density indexed by $\varphi$, which may be infinite dimensional.  We are interested in identifying $\varphi$ from the observed data distribution, which is captured by $p(x;\varphi)$ and $p(y,R=1\mid x;\varphi)$. More formally, the parameter $\varphi$ is said to be identified from the observed data, if there exists a one-to-one mapping between $\varphi$ and $\{p(x;\varphi),p(y,R=1\mid x;\varphi)\}$. It is well known that $\varphi$ cannot be identified from the observed data in the absence of further assumptions \citep{robins1997toward,10.1007/978-1-4612-1284-3_1, laan2003unified}.   In this paper, we adopt the  instrumental variable framework by assuming that  $Z$ should affect nonresponse, but not the outcome, within strata of measured baseline covariates. 
\begin{assumption}
\label{assp:ind}
$Z\not\!\perp\!\!\!\perp R\mid U$ (instrumental variable relevance). $Z\perp Y\mid U$ (exclusion restriction). 
\end{assumption}

Assumption \ref{assp:ind} is consistent with the identifiability conditions in prior works with instrumental variables \citep{heckman1974shadow,heckman1979sample,manski1990nonparametric,ahn1993semiparametric,powell1994estimation, das2003nonparametric,tchetgen2017general,sun2018semiparametric}. The exclusion restriction implies that the model for the full data density can be factorized as $p(w;\psi,\beta,\zeta)=p(y\mid u;\psi)p(z\mid u;\beta)p(u;\zeta)$. Nonetheless, $\varphi$ cannot be identified from the observed data even with exclusion restriction, as illustrated by the following example.

\begin{example}
\label{ex2}
Suppose $\mathcal{Y}=\mathcal{Z}=\{0,1\}$, and there are no baseline covariates. Then we are able to identify the five parameters $p(y,R=1 \mid z)$, $p(Z=1)$ from the observed data, but under assumption \ref{assp:ind}, there are six unknown parameters $p(Y=1)$, $p(Z=1)$ and $p(R=1 \mid y,z)$, which remains unidentifiable by parameter counting.
\end{example}

Therefore, we need to restrict the candidates for the complete data distribution to an even smaller set. Specifically, a working parametric model is assumed for the selection bias function
\begin{align}
\label{eq:sb}
h(y,x)=\log\left\{\frac{p(R=1\mid y,x)/p(R=0\mid y,x)}{p(R=1\mid Y=0,x)/p(R=0\mid Y=0,x)}\right\},
\end{align}
which encodes our {\it a priori} belief of the way in which the  underlying outcome affects the response mechanism on the log odds ratio scale. This implies the following semiparametric exponential tilting model for the propensity score,
\begin{align}
\label{eq:set}
\mathcal{}\pi(w;\eta,\gamma)=\textup{expit}\{\eta(x)+h(y,x;\gamma)\},
\end{align}
where $\textup{expit}(t)=1/\{1+\exp(-t)\}$ is the inverse logit function, $\eta(x)$ is an unknown function, and $h(y,x;\gamma)$  is a known function smooth in the finite-dimensional parameter $\gamma\in\mathbb{R}^{p_{\gamma}}$, which satisfies $h(0,x;\gamma)=0$ based on the odds ratio representation (\ref{eq:sb}). In addition, the parameterization is typically chosen to be such that $h(y,x;0)=0$,  so that $\gamma=0$ corresponds to the ignorable nonresponse mechanism. A common specification is $h(y,x;\gamma)=\gamma y$ \citep{kim2011semiparametric,shao2016semiparametric}, and in general the selection bias function can be specified more flexibly to allow for dependence of such effects on fully observed covariates.  

The missingness is nonignorable if and only if the true parameter value $\gamma_0\neq 0$. As  $\gamma_0$ is not identified in the absence of further assumptions, sensitivity analysis has been  proposed whereby one conducts inferences assuming $\gamma=\gamma_0$ is completely known and repeats the analysis upon varying the assumed value of $\gamma$  \citep{rotnitzky1998semiparametric,scharfstein1999adjusting, 10.1007/978-1-4612-1284-3_1,vansteelandt2007estimation}. \citet{kim2011semiparametric} assumed that $\gamma_0$ is known or can be estimated using external data, while \citet{miao2015identification} and  \citet{shao2016semiparametric}  used shadow variables to estimate $\gamma_0$. In this paper, we follow the instrumental variable approach of \cite{sun2018semiparametric} and assume that the following condition holds for identification.
\begin{assumption}
\label{assp:selection}
The semiparametric exponential tilting model (\ref{eq:set}) is correctly specified. Furthermore, for any given $u\in\mathcal{U}$, the ratio $\pi(w;\eta_1,\gamma_1)/\pi(w;\eta_2,\gamma_2)$ is either a constant or varies with $z$ for any two values $(\eta_1,\gamma_1)$ and $(\eta_2,\gamma_2)$ of $(\eta,\gamma)$.
\end{assumption}
}
 { Note that assumption \ref{assp:selection} does not impose further restrictions on the full data density  model $p(w;\psi,\beta,\zeta)$. In the supplementary material, we show that assumptions \ref{assp:pos}--\ref{assp:selection} are sufficient for identification of the complete data distribution from the observed data.}

\subsection{Prior works}
\label{sec:pw}
We briefly review existing methods in the nonignorable missing data literature under semiparametric exponential tilting model (\ref{eq:set}). If $\gamma=\gamma_0$ is known, then the outcome mean $\mu_0=E(Y)$ can be estimated using either the inverse propensity weighting or regression approach \citep{scharfstein1999adjusting,vansteelandt2007estimation,kim2011semiparametric}, based on the representations
$$\mu_0=\mathbb{E}\left\{\frac{RY}{\pi(W)}\right\},$$
or
$$\mu_0=\mathbb{E}\{RY+(1-R)\mathbb{E}(Y\mid R=0,X)\},$$
respectively.  For the  inverse propensity weighting approach, the unknown function $\eta(x)$ is nonparametrically identified based on the conditional moment restriction,
\begin{align}
\label{eq:exp}
\mathbb{E}\left\{\frac{R}{\pi(W;\eta,\gamma)}-1\biggr\rvert X \right\}=0,
\end{align}
for each fixed value of $\gamma$ \citep{vansteelandt2007estimation,kim2011semiparametric, shao2016semiparametric}. For the second representation, the odds ratio factorization of the complete data distribution is  widely adopted, in which the density among nonrespondents can be expressed as an exponential tilting of the density among respondents, 
\begin{align}
\label{eq:exp2}
p(y\mid R=0,x;\gamma)=\frac{\exp\{-h(y,x;\gamma)\}p(y\mid R=1,x)}{\int_{\mathcal{Y}} \exp\{-h(t,x;\gamma)\}p(t\mid R=1,x)d\nu(t)},
\end{align}
where $\nu$ is an appropriate dominating measure \citep{yun2007semiparametric,vansteelandt2007estimation,tchetgen2010doubly,kim2011semiparametric,riddles2016propensity}. Thus, if $\gamma=\gamma_0$ is known or can be estimated using external data, then estimation of $\mu_0$ entails estimation of either $\eta(x)$ or $p(y\mid R=1,x)$. When $X$ is high-dimensional or contains numerous continuous components, we can specify parametric or semiparametric models for these unknown functions. In the absence of further restrictions, the parameterization of the models for $\eta(x)$ and $p(y\mid R=1,x)$ are variationally independent \citep{yun2007semiparametric}. This provides the basis for doubly robust inference about the outcome mean if either the model for $\eta(x)$ or $p(y\mid R=1,x)$ is correctly specified \citep{vansteelandt2007estimation}.  {In particular, the value $\gamma=0$ corresponds to doubly robust inference under ignorable nonresponse.}

If the true value $\gamma_0$ is unknown,  \citet{miao2015identification}, \citet{miao2016varieties} and \citet{sun2018semiparametric} developed doubly robust inference for $\mu_0$ under a similar odds ratio factorization of the complete data distribution. However, the possibility for genuine doubly robust inference with instrumental variables is complicated by the fact that the models for $\eta(x)$ and $p(y\mid R=1,x)$  are variationally dependent under the exclusion restriction of assumption \ref{assp:ind}. To characterize this dependency, the propensity score $\tilde{\pi}(x)=p(R=1\mid x)$ can be expressed under model (\ref{eq:set})  as $\tilde{\pi}(x)=1/\{1+\iota(x)\}$, where $\iota(x)=\mathbb{E}[\exp\{-\eta(X)-h(Y,X;\gamma)\}\mid R=1,X=x]$ \citep[Lemma 8.1]{kim2021statistical}. The outcome density conditional on fully observed values is given by
\begin{align}
\label{eq:res}
p(y\mid x)= 
p(y\mid R=1,x)\tilde{\pi}(x)+p(y\mid R=0,x)\{1-\tilde{\pi}(x)\}.
\end{align}
Assumption \ref{assp:ind} imposes the additional exclusion restriction that $p(y\mid z,u)=p(y\mid u)$ for all $w\in \mathcal{W}$, which induces variation dependency between the models for $\eta(x)$ and $p(y\mid R=1,x)$.

\subsection{Alternative factorization}

In this paper, we develop novel estimators of $\mu_0$ in the semiparametric model $\mathcal{M}$ defined by assumptions  \ref{assp:pos}--\ref{assp:selection},
\begin{align*}
p(r,w;\varphi)=\pi(w;\eta,\gamma)^r\{1-\pi(w;\eta,\gamma)\}^{1-r}\underbrace{p(y\mid u;\psi)p(z\mid u;\beta)p(u;\zeta)}_{p(w;\psi,\beta,\zeta)},
\end{align*}
where $\gamma$ is finite-dimensional and $(\eta,\psi,\beta,\zeta)$ is infinite-dimensional. The full data density is clearly separated from the nonresponse mechanism in the factorization above. This naturally encodes the exclusion restriction of assumption \ref{assp:ind}, which operates on the full data distribution, rather than on subpopulations defined by the nonresponse status. Furthermore, $\psi$ and $\beta$ are variationally independent, in the sense that any appropriate choice of $\psi$ and $\beta$ would result in a density in $\mathcal{M}$. Therefore, the models $p(y\mid u;\psi)$ and $p(z\mid u;\beta)$ can  be specified separately without concerns about incompatibility. We show in the paper that estimation of  $\mu_0$  requires consistent estimation of at least a subset of the nuisance parameters $(\eta,\psi,\beta)$. In modern studies, a broad collection of baseline covariates and operational features are usually recorded. When $X$ is high-dimensional or contains several continuous components, nonparametric estimation is typically infeasible in the moderate sample sizes that are found in practice, as the data are too sparse due to the curse of dimensionality  \citep{robins1997toward}.  Thus, we are often forced to specify more stringent dimension-reducing models $\eta(x;\xi)$,  $p(y\mid u;\psi)$ and $p(z\mid u;\beta)$. Although in principle these models could be made as flexible as allowed by the sample size, we focus on parametric specifications in this paper to ease exposition. 
 
To mitigate the effects of model misspecifications, we develop novel semiparametric estimators of $\mu_0$ which remain  consistent and asymptotically normal in a union model, where either one, but not necessarily both, of the following modeling assumptions hold: 
\begin{enumerate}
\item[(1)]The models $\eta(x;\xi)$ and $p(z\mid u;\beta)$ are correctly specified such that $\eta(x)=\eta(x;\xi_0)$ and $p(z\mid u)$=$p(z\mid u;\beta_0)$    for some unknown finite-dimensional  parameter vectors $\xi_0$ and $\beta_0$;
\item[(2)]  The models $\eta(x;\xi)$ and $p(y\mid u;\psi)$ are correctly specified such that $\eta(x)=\eta(x;\xi_0)$ and $p(y\mid u)=p(y\mid u;\psi_0)$   for some unknown finite-dimensional  parameter vectors $\xi_0$ and $\psi_0$;
\end{enumerate}
Accordingly, we define the submodels $\mathcal{M}_1$, $\mathcal{M}_2$ of $\mathcal{M}$ which correspond to models (1), (2) respectively. Thus, the proposed estimators are doubly robust, as they deliver valid inferences in the union model $\cup_{j=1,2}\mathcal{M}_j$ where $\eta(x;\xi)$ (and hence the parametric extended propensity score model) is correctly specified, and either $p(z\mid u;\beta)$ or $p(y\mid u;\psi)$ is correctly specified. The proposed methodology requires correct specification of the model for $\eta(x)$, and therefore differs from the typical doubly robust inference when $\gamma=\gamma_0$ is known, where either the model for $\eta(x)$ or $p(y\mid R=1,x)$, but not necessarily both, is correctly specified. It also differs from the doubly robust inference with instrumental variables proposed by  \citet{sun2018semiparametric} when  $\gamma_0$ is unknown, where the model for $p(z\mid u)$ is correctly specified, and either the model for $\eta(x)$ or $p(y\mid R=1,x)$  is correctly specified. Specifying separate models for $\eta(x)$ and $p(y\mid R=1,x)$ in a way that respects exclusion restriction is difficult, as they are inextricably entwined in the conditional density $p(y\mid x)$. On the other hand, because $\psi$ and $\beta$ are variationally independent under exclusion restriction, the proposed methodology provides the analyst with two genuine independent opportunities to obtain valid inference.

\subsection{Notation}
Throughout, we use $A^\T$ to denote the transpose of a vector or matrix $A$, and  $A_1\otimes A_2$ to denote the Kronecker product  of two vectors or matrices $A_1$ and $A_2$. Estimation and inference are based on an independent and identically distributed observed data sample $(O_1,...,O_n)$, where $O=(R,RY,X)$. We denote the empirical measure as $\mathbb{P}_n$ so that empirical averages may be written as $n^{-1}\sum^n_{i=1}\{m(O_i)\}=\mathbb{P}_n\{m(O)\}$.  For an arbitrary function $f(W)$ of the full data $W$, denote $f^{\dag}(W)=\mathbb{E}\{f(W)\mid X\}+\mathbb{E}\{f(W)\mid Y,U\}-\mathbb{E}\{f(W)\mid U\}$.

{
 \subsection{An example with binary outcome and instrumental variable}
\label{sec:example}

We illustrate the key ideas with binary outcome and instrumental variable, $\mathcal{Y}=\mathcal{Z}=\{0,1\}$. In this case, the exclusion restriction in assumption \ref{assp:ind} is equivalent to $\textup{Cov}(Y,Z\mid U)=0$, which can be captured by the moment restriction $\mathbb{E}[Y\{Z-p(Z=1\mid U)\}\mid U]=0$. The main idea is that we can replace the moment function with its inverse propensity weighted form, to create the following  observed data conditional moment restriction for the tilting parameter $\gamma$,
\begin{align}
\label{eq:restriction1}
\mathbb{E}\left[\frac{R}{\pi(W;\eta,\gamma)}Y\{Z-p(Z=1\mid U)\}\biggr\rvert U\right]=0,
\end{align}
which implies the following unconditional form,
\begin{align}
\label{eq:moment1}
\mathbb{E}\left[\frac{R}{\pi(W;\eta,\gamma)}{d}(U)Y\{Z-p(Z=1\mid U)\}\right]=0,
\end{align}
where ${d}(u)\in\mathbb{R}^{p_{\gamma}}$ is a user-specified,  vector-valued function with linearly independent elements. In principle,  we can construct a semiparametric two-step estimator $\hat{\gamma}$ of $\gamma_0$ which solves the following empirical analogue of (\ref{eq:moment1}), $$\mathbb{P}_n\left[\frac{R}{\pi(W;\hat{\eta},\gamma)}{d}(U)Y\{Z-\hat{p}(Z=1\mid U)\}\right]=0,$$ where $\hat{\eta}(x)$ is a nonparametric  estimator of $\eta(x)$ based on (\ref{eq:exp}) for each fixed value of $\gamma$, and $\hat{p}(Z=1\mid u)$  is a nonparametric  estimator of $p(Z=1\mid u)$. We can then construct the following \citet{hajek1971comment} estimator of the outcome mean, $$\hat{\mu}={\mathbb{P}_n\left\{\frac{RY}{\pi(W;\hat{\eta},\hat{\gamma})}\right\}}\biggr\slash{\mathbb{P}_n\left\{\frac{R}{\pi(W;\hat{\eta},\hat{\gamma})}\right\}}.$$

On the other hand, the zero conditional covariance restriction under assumption \ref{assp:ind} can also be captured by the  moment restriction $\mathbb{E}[\{Y-p(Y=1\mid U)\}Z\mid U]=0$. This yields the observed data conditional moment restriction
\begin{align}
\label{eq:restriction2}
\mathbb{E}\left[\frac{R}{\pi(W;\eta,\gamma)}\{Y-p(Y=1\mid U)\}Z\biggr\rvert U\right]=0.
\end{align}
In contrast to the previous approach, $p(Y=1\mid u)$ cannot be directly estimated from observed data, but instead can be implicitly defined as the solution to the following inverse propensity weighted moment restriction,
\begin{align}
\label{eq:alpha}
\mathbb{E}\left[\frac{R}{\pi(W;\eta,\gamma)}\{Y-p(Y=1\mid U)\}\biggr\rvert U\right]=0.
\end{align}
A key observation is that (\ref{eq:restriction2}) and (\ref{eq:alpha}) are functionally different. We can then construct a semiparametric two-step estimator of $\gamma_0$ which solves the following empirical moment condition for some vector-valued function ${d}(u)\in\mathbb{R}^{p_{\gamma}}$,
\begin{align*}
\mathbb{P}_n\left[\frac{R}{\pi(W;\hat{\eta},\gamma)}{d}(U)\{Y-\hat{p}(Y=1\mid U)\}Z\right]=0,
\end{align*}
where   $\hat{\eta}(x)$ is a nonparametric estimator of $\eta(x)$ based on (\ref{eq:exp}), and $\hat{p}(Y=1\mid u)$ is a nonparametric estimator of $p(Y=1\mid u)$ based on (\ref{eq:alpha}), for each fixed value of $\gamma$. A \citet{hajek1971comment} estimator of the outcome mean can be constructed similarly.

The preceding two approaches require estimation of $\eta(x)$, and either  $p(z\mid u)$ or $p(y\mid u)$. In the presence of possibly high-dimensional $X$, we can perform inference  under working parametric specifications $\eta(x;\xi)$, $p(z\mid u;\beta)$ and  $p(y\mid u;\psi)$. We obtain $\hat{\theta}=(\hat{\xi}^\T,\hat{\beta}^\T,\hat{\psi}^\T)^\T$, for each fixed value of $\gamma$, as the solution to the following unconditional empirical moments,
\begin{align}
\label{eq:moment}
\mathbb{P}_n\left[\left\{\frac{R}{\pi(W;\xi,\gamma)}-1\right\}\frac{\partial \eta(X;\xi)}{\partial \xi}\right]=0;\nonumber \\ 
\mathbb{P}_n\left\{{\frac{\partial \log p(Z\mid U;\beta)}{\partial \beta}}\right\}=0; \\ \nonumber
\mathbb{P}_n\left[\left\{\frac{R}{\pi(W;\xi,\gamma)}\right\}{\frac{\partial \log p(Y\mid U;\psi)}{\partial \psi}}\right]=0. 
\end{align}
}

\section{Doubly Robust Inference}
\label{sec:dr}
\subsection{Semiparametric theory}

The construction of the \citet{hajek1971comment} estimators of $\mu_0$ in the previous section involves preliminary estimation of the tilting parameter $\gamma_0$. As $\gamma_0$ encodes the degree of departure from ignorability, it is also sometimes of interest in its own right in missing data problems. To simplify the presentation, in this section we consider joint estimation of the unknown $(p_{\gamma}+1)$-dimensional parameter ${{\phi}}_0=(\mu_0,\gamma^\T_0)^\T$. The construction of doubly robust estimators of ${{\phi}}_0$ is often motivated by the form of the influence function in $\mathcal{M}$ \citep{robins2001comment,chernozhukov2022locally}. Specifically, any regular and asymptotically linear estimator $\hat{{\phi}}$ of ${{\phi}}_0$ in $\mathcal{M}$ satisfies
$$n^{1/2}(\hat{\phi}-\phi_0)=n^{1/2} \mathbb{P}_n\{\Psi(O;\phi_0)\}+o_p(1),$$
where the $i$-th influence function $\Psi(O_i)$ represents the influence of the $i$-th observation on the estimator \citep{pfanzagl1982contributions,bickel1993efficient,newey1994asymptotic,laan2003unified,tsiatis2007semiparametric}. The estimation theory for $\phi_0$ under general semiparametric models for the nonresponse mechanism has been previously developed \citep{10.1007/978-1-4612-1284-3_1,laan2003unified}. The result below, proved in \citet{sun2018semiparametric},  provides an application of the general theory  to model $\mathcal{M}$.

\begin{result}
\label{lem:1}
The observed data influence function of any regular and asymptotically linear estimator of $\phi_0$ in $\mathcal{M}$ is given by
\begin{align*}
\Psi(O;\phi_0,c,d)=-\left[\mathbb{E}\left\{\frac{\partial g(O;\phi,c,d)}{\partial \phi} \right\}{\biggr\rvert_{\phi={\phi}_0}}\right]^{-1} g(O;\phi_0,c,d),
\end{align*} 
for some functions $c(W)\in \mathbb{R}$ and $d(W)\in \mathbb{R}^{p_{\gamma}}$ of the full data $W$, where
$$
 g(O;\phi,c,d)=\frac{Rq(W;\mu,c,d)}{\pi(W;\gamma)}+{\left\{1-\frac{R}{\pi(W;\gamma)}\right\}\mathbb{E}\left\{
q(W;\mu,c,d)\rvert R=0,X;\gamma\right\}},
$$
and
$$
q(W;\mu,c,d)=\left\{
\begin{matrix}
Y-\mu+c(W)-c^{\dag}(W)\\
d(W)-d^{\dag}(W)
\end{matrix}\right\}\in\mathbb{R}^{p_{\gamma}+1}.
$$
\end{result}
\noindent The function $g(O;\phi,c,d)$ has the familiar inverse probability weighted form in missing data literature, augmented by an additional term involving the nonrespondents \citep{robins1994estimation,10.1007/978-1-4612-1284-3_1,laan2003unified}. 

\begin{example}

In the special case with $\mathcal{Y}=\mathcal{Z}=\{0,1\}$, any function of the full data $W$ can be expressed as $f(W)=f_0(U)+Yf_y(U)+Zf_z(U)+YZ f_{yz}(U)$, where $\{f_0(U),f_y(U),f_z(U),f_{yz}(U)\}$ are functions of the same dimension as $f(W)$. It can be shown that 
\begin{align*}
f(W)-f^{\dag}(W)&={f}_{yz}(U)\{Y-p(Y=1\mid U)\}\{Z-p(Z=1\mid U)\}.
\end{align*}
The example generalizes directly to discrete outcome and instrument taking values in $\mathcal{Y}=\{0,1,...,\ell_y\}$ and $\mathcal{Z}=\{0,1,...,\ell_z\}$, respectively. Let $v_1(Y)=\{I(Y=1),...,I(Y=\ell_y)\}^\T$ and $v_2(Z)=\{I(Z=1),...,I(Z=\ell_z)\}^\T$, where $I(\cdot)$ is the indicator function. Then for any function of the full data $W$,
\begin{align*}
f(W)-f^{\dag}(W)={f}_{yz}(U)\{v_1(Y)-\mathbb{E}(v_1(Y)\mid U)\}\otimes\{v_2(Z)-\mathbb{E}(v_2(Z)\mid U)\},
\end{align*}
for some conformable ${f}_{yz}(U)$. 

\end{example}

\begin{example}
If at least one of $Y$ and $Z$ is continuous, inspired by the discrete case, we can simply discretize the continuous variables or use their moments. For example, if both $Y$ and $Z$ are continuous, we can consider the vectors ${v}_1(Y)=(Y,Y^2,...,Y^{p_y})^\T$ and ${v}_2(Z)=(Z,Z^2,...,Z^{p_z})^\T$, for some positive integers $p_y$ and $p_z$.
\end{example}

\subsection{Influence function-based doubly robust estimators}

Following \citet{robins2001comment} and \citet{chernozhukov2022locally}, we can use $g(O;\phi,c,d)$ as a moment function to estimate $\phi_0$. Evaluation of $g(O;\phi,c,d)$ relies on $\{\eta(x),p(z\mid u),p(y\mid u)\}$, which are directly targeted by the proposed approach. In particular, we note that the conditional outcome density among non-respondents can be expressed as
\begin{align}
\label{eq:nonrespondents}
p(y\mid R=0,x;\gamma)=\frac{\{1-\pi(y,x;\gamma)\}p(y\mid u)}{1-\int_{\mathcal{Y}} \pi(t,x;\gamma)p(t\mid u)d\nu(t)},
\end{align}
which differs from (\ref{eq:exp2}) in terms of parameterization. When the baseline covariates include multiple continuous components, $\{\eta(x),p(z\mid u),p(y\mid u)\}$ can be estimated under user-specified, dimension-reducing parametric  specifications $\{\eta(x;\xi),p(z\mid u;\beta), p(y\mid u;\psi)\}$, which allows for simpler conditions for asymptotic normality. Let $m(O;\gamma,\theta)$ denote the stacked moment functions in (\ref{eq:moment}) for the parameter $\theta=(\xi^\T,\beta^\T,\psi^\T)^\T$. The proposed influence function-based estimator $(\hat{\phi}^\T(c,d),\hat{\theta}^\T)^\T$ then solves  $$\mathbb{P}_n\{g^\T(O;\phi,\theta,c,d),m(O;\gamma,\theta)\}^\T=0.$$ The asymptotic property of $\hat{\phi}(c,d)$ is given in the next proposition, where $\bar{\theta}=(\bar{\xi}^\T,\bar{\beta}^\T,\bar{\psi}^\T)^\T$ denotes the probability limit of $\hat{\theta}$.

\begin{prop}
\label{lem:dr1}
Under standard regularity conditions for moment estimation \citep{newey1994large}, the estimator $\hat{\phi}(c,d)$ admits the following asymptotic expansion in the union model $\cup_{j=1,2}\mathcal{M}_j$, 
\begin{align*}
n^{1/2}(\hat{\phi}(c,d)-{\phi}_0)=-n^{1/2}\left[\mathbb{E}\biggr\{\frac{\partial}{\partial \phi} G(O;\phi,\bar{\theta},c,d)\biggr\}{\biggr \rvert_{\phi={\phi}_0}}\right]^{-1}\mathbb{P}_n\{G(O;{\phi}_0,\bar{\theta},c,d)\}+o_p(1),
\end{align*}
where
\begin{align*}
G(O;\phi,\bar{\theta},c,d)=g&(O;\phi,\bar{\theta},c,d)\\
&-\mathbb{E}\biggr\{\frac{\partial}{\partial \theta} g(O;{\phi}_0,{\theta},c,d)\biggr\}{\biggr \rvert_{\theta=\bar{\theta}}}\mathbb{E}\biggr\{\frac{\partial}{\partial \theta} m(O;{\gamma}_0,{\theta})\biggr\}^{-1}{\biggr \rvert_{\theta=\bar{\theta}}}{ m(O;\gamma,\bar{\theta})}.
\end{align*}
Furthermore, at the intersection submodel $\cap_{j=1,2}\mathcal{M}_j$ where all the working parametric models are correctly specified, $\hat{\phi}(c,d)$ admits the asymptotic expansion 
\begin{align*}
n^{1/2}(\hat{\phi}(c,d)-\phi_{0})=n^{1/2}\mathbb{P}_n\{\Psi(O;\phi_0,c,d)\}+o_p(1).
\end{align*}
\end{prop}
\noindent The first part of proposition \ref{lem:dr1} is due to the following double robustness property,
\begin{align}
\label{eq:drprop}
\mathbb{E}\{{g}(O;\phi_0,\bar{\xi},\bar{\beta},\bar{\psi},c,d)\}=0,
\end{align}
if either $\{\eta(x;\bar{\xi}),p(z\mid u ;\bar{\beta})\}=\{{\eta}(x),p(z\mid u)\}$ or $\{\eta(x;\bar{\xi}),p(y\mid u ;\bar{\psi})\}=\{{\eta}(x),p(y\mid u)\}$. Thus, deviations of $p(z\mid u ;\bar{\beta})$ or $p(y\mid u ;\bar{\psi})$ away from the truth  has no {global} effect on the moment condition. The second part of proposition \ref{lem:dr1} states that estimation of the nuisance parameter $\theta$ has no first-order effect on the asymptotic expansion of $\hat{\phi}(c,d)$ at the intersection submodel $\cap_{j=1,2}\mathcal{M}_j$ where  $\{\eta(x;\bar{\xi}),p(z\mid u ;\bar{\beta}),p(y\mid u ;\bar{\psi})\}=\{{\eta}(x),p(z\mid u),p(y\mid u)\}$. This is a general property of estimators that are constructed based on influence functions due to Neyman orthogonality \citep{chernozhukov2022locally}, which allows for simplification of the asymptotic variance formula. However, in general this simplification is lost as soon as one of the working models $p(z\mid u ;{\beta})$ or $p(y\mid u ;{\psi})$ is misspecified \citep{vermeulen2015bias}.

\subsection{Local efficiency}

When both $Y$ and $Z$ are discrete, the efficient influence function in $\mathcal{M}$ is indexed by the optimal choice $(c,d)=(c^{\ast},d^{\ast})$, which is characterized in the supplementary material. At the intersection submodel $\cap_{j=1,2}\mathcal{M}_j$, the influence function of $\hat{\phi}(c,d)$ coincides with $\Psi(O;\phi_0,c,d)$, and hence the asymptotic variance of $\hat{\phi}(c^{\ast},d^{\ast})$ attains the semiparametric efficiency bound $\mathcal{V}=\mathbb{E}\{\Psi(O;\phi_0,c^{\ast},d^{\ast})\Psi(O;\phi_0,c^{\ast},d^{\ast})^\T\}$ (local efficiency). In practice, we can implement a doubly robust and locally efficient estimator of $\phi_0$ based on a preliminary, consistent estimator of the optimal index functions, which we describe in the supplementary material. The results in \citet{robins2001comment} show that the efficiency bound in the union model $\cup_{j=1,2}\mathcal{M}_j$ coincides with the semiparametric efficiency bound $\mathcal{V}$ in $\mathcal{M}$. Therefore, $\hat{\phi}(c^{\ast},d^{\ast})$ also attains the efficiency bound in $\cup_{j=1,2}\mathcal{M}_j$ at the intersection submodel $\cap_{j=1,2}\mathcal{M}_j$. When at least one of $Y$ and $Z$ is continuous, the efficient influence function is typically not available in closed form. In this case, we can construct a doubly robust and approximately locally efficient estimator of $\phi_0$, which is described in  \citet{sun2018semiparametric}.

\subsection{A computationally simpler doubly robust estimator}

The implementation of $\hat{\phi}(c,d)$ requires evaluation of the conditional outcome density among non-respondents given in (\ref{eq:nonrespondents}). It is generally difficult to evaluate the integral in the denominator. For example,  no closed form result for the  logistic-normal integral is known, although various approximations have been proposed in the literature \citep{crouch1990evaluation}. In this section, we propose a computationally simpler doubly robust estimator in the union model $\cup_{j=1,2}\mathcal{M}_j$ which avoids the numerical integration. The key observation is that double robustness (\ref{eq:drprop}) continues to hold for the inverse propensity weighted function $\tilde{g}(O;\phi,c,d)={Rq (W;\mu,c,d)}/{\pi(W;\gamma)}$ which excludes the augmentation term  involving  non-respondents. Let $(\tilde{\phi}^\T(c,d),\hat{\theta}^\T)^\T$ denote the joint solution to  $$\mathbb{P}_n\{\tilde{g}^\T(O;\phi,\theta,c,d),m(O;\gamma,\theta)\}^\T=0.$$ 

\begin{example}
To illustrate the proposed  computationally simpler doubly robust estimators, suppose $\mathcal{Y}=\mathcal{Z}=\{0,1\}$ and we set $c(W)=0$. In this case, following the examples in section \ref{sec:example}, estimation based on $\tilde{g}(O;\phi,c=0,d)$ can proceed in two stages. In the first stage, $(\tilde{\gamma}^\T,\tilde{\theta}^\T)^\T$ solves (\ref{eq:moment}) jointly with $$\mathbb{P}_n\left[\frac{R}{\pi(W;\xi,\gamma)}d_{yz}(U)\{Y-p(Y=1\mid U;\psi)\}\{Z-p(Z=1\mid U;\beta)\}\right]=0,$$
for some $d_{yz}(U)\in \mathbb{R}^{p_{\gamma}}$. Then the \citet{hajek1971comment} estimator $\tilde{\mu}$ of the outcome mean can be constructed with the estimated inverse propensity weights $\{R_i/\pi(W_i;\tilde{\xi},\tilde{\gamma}):i=1,...,n\}$. 
\end{example}

\noindent The asymptotic property of $\tilde{\phi}(c,d)$ is summarized in the next proposition.

\begin{prop}
\label{lem:dr2}
Under standard regularity conditions for moment estimation \citep{newey1994large}, the estimator $\tilde{\phi}(c,d)$ admits the following asymptotic expansion in the union model $\cup_{j=1,2}\mathcal{M}_j$, 
\begin{align*}
n^{1/2}(\tilde{\phi}(c,d)-{\phi}_0)=-n^{1/2}\left[\mathbb{E}\biggr\{\frac{\partial}{\partial \phi}  \tilde{G}(O;\phi,\bar{\theta},c,d)\biggr\}{\biggr \rvert_{\phi={\phi}_0}}\right]^{-1}\mathbb{P}_n\{ \tilde{G}(O;{\phi}_0,\bar{\theta},c,d)\}+o_p(1),
\end{align*}
where
\begin{align*}
 \tilde{G}(O;{\phi},\bar{\theta},c,d)=\tilde{g}&(O;\phi,\bar{\theta},c,d)\\
&-\mathbb{E}\biggr\{\frac{\partial}{\partial \theta} \tilde{g}(O;{\phi}_0,{\theta},c,d)\biggr\}{\biggr \rvert_{\theta=\bar{\theta}}}\mathbb{E}\biggr\{\frac{\partial}{\partial \theta} m(O;{\gamma}_0,{\theta})\biggr\}^{-1}{\biggr \rvert_{\theta=\bar{\theta}}}{ m(O;\gamma,\bar{\theta})}.
\end{align*}
\end{prop}
The asymptotic variance of $\tilde{\phi}(c,d)$ cannot attain the efficiency bound in $\mathcal{M}$ (and hence also in $\cup_{j=1,2}\mathcal{M}_j$) even at the intersection submodel $\cap_{j=1,2}\mathcal{M}_j$. Therefore, the estimator $\tilde{\phi}(c,d)$ is doubly robust but not locally efficient, and intuitively this is because it fails to  incorporate the information from non-respondents when all the working models are correctly specified. Nonetheless, because we have  already paid homage to the need for efficiency by using parametric models, the potential prize of attempting to attain local efficiency may not always be worth the chase in view of the additional computational demands.  As the goal of this paper is to produce a statistically sound and practically useful method to tackle nonignorable missing data, we will focus on the estimator $\tilde{\phi}(c,d)$ in the simulation study and application.

\section{Simulation Studies}
\label{sec:sim}

In order to investigate the finite-sample properties of the doubly robust estimator proposed in Section \ref{sec:dr}, we perform Monte Carlo simulations involving identical and independently generated data $\{O_1,...,O_n\}$. The baseline covariate $U=(U_1,U_2)^\T$ is generated from a bivariate normal distribution $N(0,\Sigma)$, where the elements of $\Sigma$ are $\sigma^2_1=\sigma^2_2=1$ and $\sigma_{12}=0.2$. Conditional on $U$, $(R,Y,Z)$ is generated from the following  generalized linear models consistent with assumptions  \ref{assp:pos}--\ref{assp:selection}, 
 \begin{align*}
Z\mid U &\sim \textup{Bernoulli}\{p_1=\textup{expit}(1+2U_1-U_2-0.8U_1U_2)\},\\
Y\mid Z,U &\sim \textup{Bernoulli}\{p_2=\textup{expit}(0.5-2U_1+U_2)\},\\
R\mid Y,X &\sim \textup{Bernoulli}\{\pi=\textup{expit}(2-3Z+0.8U_1+U_2+\gamma Y)\},
 \end{align*}
 where $\gamma=2$. { We implement the proposed doubly robust estimator $\tilde{\phi}(c,d)=(\tilde{\mu},\tilde{\gamma})^\T$ with $c(w)=0$, $d(w)=yz$ based on the models $\pi(w;\xi,\gamma)=\textup{expit}\{(h^\T_1(u),z)^\T\xi+\gamma y\}$, $p(Z=1\mid u;\beta)=\textup{expit}\{h_2(u)\beta\}$ and $p(Y=1\mid u;\psi)=\textup{expit}\{h_3(u)\psi\}$ under the following five scenarios. The model for $\eta(x)$, $p(Z=1\mid u)$ or $p(Y=1\mid u)$ is misspecified if $h_j(u)=(1,u_1,u^2_1)$ for $j=1$, 2 or 3, respectively.
\begin{itemize}
\item[(C1)]  Models for $\{\eta(x),p(Z=1\mid u),p(Y=1\mid u)\}$ are all correct.
\item[(C2)] Models for $\{\eta(x), p(Y=1\mid u)\}$ are correct, but misspecified for $p(Z=1\mid u)$.
\item[(C3)] Models for $\{\eta(x), p(Z=1\mid u)\}$ are correct, but misspecified for $p(Y=1\mid u)$.
\item[(C4)] Models for $\{p(Z=1\mid u),p(Y=1\mid u)\}$ are correct, but misspecified for $\eta(x)$.
\item[(C5)]  Models for $\{\eta(x),p(Z=1\mid u),p(Y=1\mid u)\}$ are all misspecified.
\end{itemize}
To compare $\tilde{\mu}$ with the doubly robust estimator $\hat{\mu}_{{\textup{dr}}}$ proposed by \citet{sun2018semiparametric}, we specify the working parametric model $p(Y=1\mid R=1, x;\psi_1,\lambda)=\textup{expit}\{ h_3(u)\psi_1+\lambda z \}$ under (C1)--(C5) to evaluate its use in practice. In addition, we implement the complete-case estimator of the outcome mean $\hat{\mu}_{{\textup{cc}}}=\mathbb{P}_n(RY)$, as well as the infeasible full-data estimator $\hat{\mu}_{{\text{full}}}=n^{-1}\sum^{n}_{i=1}Y_i $  as  performance benchmark. For inference, we construct 95\% Wald confidence intervals based on the sandwich estimator of asymptotic variance.

{ The following remarks can be made based on the results of 1000 simulation replicates of sample size $n=500$, $1000$  or $5000$ summarized in Table \ref{tab:sim}.}  The complete-case estimator $\hat{\mu}_{{\textup{cc}}}$ exhibits severe bias and undercoverage. In agreement with theory, $\tilde{\mu}$ performs well in terms of bias and coverage cross scenarios (C1)--(C3), but is biased under (C4) and (C5)  with misspecified model for $\eta(x)$. The estimator  $\hat{\mu}_{{\textup{dr}}}$ shows negligible bias and coverage proportion close to the nominal level under scenarios (C1) and (C3), but is biased under (C2) and (C5) with misspecified model for $p(Z=1\mid u)$. It also has  small but noticeable bias under (C4). The relative efficiency of  $\tilde{\mu}$ compared to the full-data estimator $\hat{\mu}_{{\textup{full}}}$ is approximately $0.15$ based on Monte Carlo variance at $n=5000$. The supplementary material contains  additional Monte Carlo simulation results  under violations of the exclusion restriction in assumption \ref{assp:ind}, as well as with a continuous $Y$.

\begin{table}
\begin{center}
  \begin{threeparttable}

		\small
		\caption{Summary of results for estimation of the outcome mean. }
		  \label{tab:sim}
		\bigskip
		\begin{tabular}{ccccccccccccc}
			\toprule
       &     &  &\multicolumn{2}{c}{(C1)} & \multicolumn{2}{c}{(C2)}& \multicolumn{2}{c}{(C3)} &\multicolumn{2}{c}{(C4)}&\multicolumn{2}{c}{(C5)}\\
        &     &             &\multicolumn{2}{c}{All correct} & \multicolumn{2}{c}{mis $p(z\mid u)$}& \multicolumn{2}{c}{mis $p(y\mid u)$} &\multicolumn{2}{c}{mis $\eta(x)$}&\multicolumn{2}{c}{All mis}\\
&	$\hat{\mu}_{{\textup{cc}}}$ &$\hat{\mu}_{{\textup{full}}}$     &  $\tilde{\mu}$ & $\hat{\mu}_{{\textup{dr}}}$ & $\tilde{\mu}$ & $\hat{\mu}_{{\textup{dr}}}$  & $\tilde{\mu}$ & $\hat{\mu}_{{\textup{dr}}}$ & $\tilde{\mu}$ & $\hat{\mu}_{{\textup{dr}}}$& $\tilde{\mu}$ & $\hat{\mu}_{{\textup{dr}}}$ \\
	     \hline
&  \multicolumn{12}{c}{$n=500^\dag$}\\
$|\text{Bias}|$ &.136 & .001 & .004 & .004 & .002 & .089 & .006 & .004 & .042 & .032 & .093 & .103\\
$\sqrt{\text{Var}}$ & .024 & .021 & .055 & .050 & .055 & .046 & .057 & .050 & .049 & .097 & .047 & .040 \\
$\sqrt{\text{EVar}}$ &.026 & .022 & .055 & .080 & .056 & 6.349 & .064 & .070 & .059 & .042 & .048 & .045\\
Cov95 & .000& .957 & .912 & .943 & .920 & .948 & .901 & .943 & .848 & .847 & .499 & .386 \\\cline{2-13}
&  \multicolumn{12}{c}{$n=1000$}\\
$|\text{Bias}|$ &.136 & .000 & .004 & .001 & .002 & .088 & .005 & .001 & .042 & .017 & .092 & .101\\
$\sqrt{\text{Var}}$ & .017 & .016 & .039 & .036 & .039 & .032 & .041 & .036 & .034 & .034 & .033 & .028\\
$\sqrt{\text{EVar}}$ & .018 & .016 & .037 & .039 & .038 & 1.813 & .039 & .039 & .034 & .029 & .034 & .032\\
Cov95 & .000 & .947 & .916 & .929 & .931 & .939 & .920 & .941 & .750 & .862 & .219 & .084\\\cline{2-13}
&  \multicolumn{12}{c}{$n=5000$}\\
$|\text{Bias}|$ &.136  & .000& .001 & .000 & .001& .090 & .002 & .000 & .044 & .013 & .093 & .102\\
$\sqrt{\text{Var}}$ & .008 & .007 & .017 & .015 & .017 & .014 & .018 & .015 & .015 & .015 & .015 & .013\\
$\sqrt{\text{EVar}}$ & .008 & .007 & .017 & .016 & .017 & .066 & .018 & .016 & .015 & .013 & .015 & .014 \\
Cov95 & .000& .954 & .939 & .950 & .943 & .578 & .933 & .951 & .169 & .806 & .000 & .000\\\hline
		\end{tabular}
   \begin{tablenotes}
      \item  {\noindent\footnotesize Note: $^\dag$The results for $\hat{\mu}_{{\textup{dr}}}$ excluded five simulation replicates due to convergence failure at $n=500$. $|\text{Bias}|$ and $\sqrt{\text{Var}}$ are the Monte Carlo absolute bias and standard deviation of the point estimates,  $\sqrt{\text{EVar}}$ is the
square root of the mean of the variance estimates and Cov95 is the coverage proportion of
the 95\% confidence intervals, based on 1000 repeated simulations.  Zeros denote values smaller than $.0005$.  The semiparametric efficiency bound for estimation of the outcome mean under the data generating mechanism of the simulation study is $\mathcal{V}\approx 1.2$ by Monte Carlo integration, with $\sqrt{\mathcal{V}/n}=.049, .035$ and $.015$ for $n=500$, 1000 and 5000 respectively.

}
    \end{tablenotes}

  \end{threeparttable}
	\end{center}
  \end{table}
	
}

\section{Illustration}
\label{sec:illus}

To illustrate the proposed method, we  analyze the household survey data on $n=4997$ adults between the ages of 16 and 64  in Mochudi, Botswana, out of whom 4045 (81\%) had complete information on HIV testing. The majority of those who did not have HIV test results refused to participate in the HIV testing component. The baseline covariates  include participant gender $(U_1)$ and  age in years $(U_2)$, while the candidate outcome instrument $Z$ is a binary indicator of whether interviewer experience in years is in the top quartile. We implement the proposed estimator  $\tilde{\phi}(c,d)=(\tilde{\mu},\tilde{\gamma})^\T$ with $c(w)=0$, $d(w)=yz$ based on the following  main effects generalized linear models with canonical links, $\pi(w;\xi,\gamma)=\textup{expit}\{(1,z,u_1,u_2)\xi+\gamma y\}$, $p(Z=1\mid u;\beta)=\textup{expit}\{(1,u_1,u_2)\beta\}$, and $p(Y=1\mid u;\psi)=\textup{expit}\{(1,u_1,u_2)\psi\}$. These parametric models are chosen due to their simplicity for illustration, although in principle they can be checked against the observed data using  goodness-of-fit tests. 

{ For comparison, we also implement the estimator $\hat{\mu}_{{\textup{dr}}}$ of \citet{sun2018semiparametric} based on the working model $p(Y=1\mid R=1, x;\psi_1,\lambda)=\textup{expit}\{ (1,u_1,u_2)\psi_1+\lambda z \}$, as well as the standard complete-case estimator $\hat{\mu}_{{\textup{cc}}}$ and the inverse propensity weighted estimator $\hat{\mu}_{{\textup{mar}}}$ based on the missing at random propensity score model $\pi(w;\xi,\gamma=0)=\textup{expit}\{(1,z,u_1,u_2)\xi\}$. The  analysis results  are summarized in Table \ref{tab:2}. The point estimates of  HIV seroprevalence for $\tilde{\mu}$ and $\hat{\mu}_{{\textup{dr}}}$ are similar and substantially higher than those for $\hat{\mu}_{{\textup{cc}}}$ and $\hat{\mu}_{{\textup{mar}}}$,  although at the expense of higher variance. This difference in efficiency reflects genuine uncertainty about the underlying nonresponse mechanism, because $\hat{\mu}_{{\textup{cc}}}$ and $\hat{\mu}_{{\textup{mar}}}$ impose {\it a priori} restrictions on the parameter space of $(\xi^\T,\gamma)^\T$. The point estimate of $\tilde{\gamma}$ is $-1.854$ with  95\% confidence interval $(-5.984,2.277)$, which suggests that HIV-infected persons are less likely to participate in the HIV testing component of the survey, although this difference is not statistically significant at the 0.05 significance level.

\begin{table}
\begin{center}
  \begin{threeparttable}
 \small
\caption{Estimation of HIV prevalence   based on a household survey data among adults in Mochudi, Botswana.}
\label{tab:2}
{%
\begin{tabular}{ccccccccc}
 \\
$\hat{\mu}_{{\textup{cc}}}$ &$\hat{\mu}_{{\textup{mar}}}$ &$\hat{\mu}_{{\textup{dr}}}$ & $\tilde{\mu}$\\
\hline
  .214 &  .213  & .284& .283\\
   $(.202,.227)$ &  $(.200,.225)$ & $(.119,.449)$ &$(.119, .447)$\\
   \hline
\end{tabular}}
   \begin{tablenotes}
      \item  {\noindent\footnotesize Note: Point estimates and 95\% confidence intervals in parenthesis.
}
    \end{tablenotes}

  \end{threeparttable}
	\end{center}

\end{table}

}

\section{Extension to longitudinal studies with repeated outcome measures}
\label{sec:extension}

We follow the longitudinal study design in \cite{vansteelandt2007estimation},  in which the full data at each study cycle $1\leq t\leq T$ consists of $W_t=(Y_t,{Z_t,U_t})$, where $Y_t$ is an outcome, $Z_t$ an instrumental variable such as interviewer experience, and $U_t$ consists of other measured covariates. Suppose $X_t=(Z_t,U_t)$ is fully observed, but $Y_t$ is subject to missingness due to some subjects missing some study cycles. Let $R_t$ denote the occasion-specific missingness status. Thus, we only observe $O_t=(R_t,R_t Y_t,X_t)$ at study cycle $t$. Here the nonresponse patterns can be nonmonotone, as $R_t = 0$ does not necessarily  imply that $R_{t+1} = 0$.  Let $\bar{O}_t=(O_1,...,O_t)$ denote the observed history up to and including cycle $t$, with $\bar{O}_0=\emptyset$. The nonresponse mechanism at each study cycle $t$ is captured by the occasion-specific extended propensity score $\pi_{t}(w_t,\bar{o}_{t-1})=p(R_t=1\mid W_t=w_t,\bar{O}_{t-1}=\bar{o}_{t-1})$. 

Suppose our interest lie in estimating the mean of the outcome vector $Y=(Y_1,...,Y_T)^\T$, which we denote by $\mu_0=(\mu_{1,0},...,\mu_{T,0})^\T$. If the interviewers in the follow-up study are deployed randomly at the start of each cycle $t$, possibly within strata of all hitherto observed variables $(U_t,\bar{O}_{t-1})$, then it is plausible that the instrumental variable conditions $Z_t\not\!\perp\!\!\!\perp R_t\mid (U_t,\bar{O}_{t-1})$ and $Z_t\perp Y_t\mid (U_t,\bar{O}_{t-1})$ hold at each study cycle $t$. It follows that, analogous to the development in section \ref{sec:dr}, an occasion-specific  doubly robust estimator of $\mu_{t,0}$ can be constructed for all $1\leq t\leq T$. The resulting estimator of $\mu_0$ is $2^T$-multiply robust \citep{vansteelandt2007estimation}, as it remains consistent so long as one of two sets of parametric models  is correctly specified at each study cycle $t$. This robustness against model misspecification is appealing in view of the possibly high-dimensional data cumulatively observed over the study cycles.

\section{Discussion}
\label{sec:discussion}

In closing, we acknowledge certain limitations of the proposed semiparametric method and outline avenues for future work.  The proposed doubly robust estimators can be improved in terms of efficiency \citep{tan2006distributional,tan2010bounded}, bias \citep{vermeulen2015bias} and robustness \citep{han2013estimation,li2020demystifying}.   Our framework also opens
the way to using various flexible  nonparametric or machine learning methods to estimate the nuisance parameters \citep{Chernozhukov2018ddml,chernozhukov2022locally}. {Although we have used the odds ratio, the proposed framework can in principle be extended to other measures of the residual association between the outcome of interest and nonresponse mechanism, for example on the multiplicative or additive scales.} Lastly, nonignorable missing covariate data is also a long-standing problem in applied research, and methods for identification and inference abound. For example,  \citet{miao2018identification} used shadow variables, while \citet{bartlett2014improving} and \citet{yang2019causal} assume that the covariate missingness mechanism is independent of the outcome. It will be of interest to explore the use of instrumental variables in similar settings with missing covariate data.


\section*{Supplementary Materials}

The online Supplementary Material contains proofs of propositions \ref{lem:dr1} and \ref{lem:dr2}, a further discussion on local efficiency and additional simulation results.

\par
\section*{Acknowledgements}

Baoluo Sun's work is supported by the Ministry of Education, Singapore, under its Academic Research Fund Tier 1 (A-8000452-00-00).  The authors would like to thank the anonymous referees, an Associate Editor
and the Editor for their constructive comments that led to a much improved paper.

\par


\bibhang=1.7pc
\bibsep=2pt
\fontsize{9}{14pt plus.8pt minus .6pt}\selectfont
\renewcommand\bibname{\large \bf References}
\expandafter\ifx\csname
natexlab\endcsname\relax\def\natexlab#1{#1}\fi
\expandafter\ifx\csname url\endcsname\relax
  \def\url#1{\texttt{#1}}\fi
\expandafter\ifx\csname urlprefix\endcsname\relax\def\urlprefix{URL}\fi

\bibliographystyle{chicago}      
\bibliography{bibfile}   

\vskip .65cm
\noindent
Department of Statistics and Data Science, National University of Singapore, Singapore.
\vskip 2pt
\noindent
E-mail: stasb@nus.edu.sg
\vskip 2pt

\noindent
Department of Probability and Statistics, Peking University, China.
\vskip 2pt
\noindent
E-mail: mwfy@pku.edu.cn
\vskip 2pt

\noindent
Department of Statistics and Data Science, National University of Singapore, Singapore.
\vskip 2pt
\noindent
E-mail: deshanee.w@u.nus.edu

\end{document}